\documentclass[a4paper,11pt]{article}
\usepackage[german,american]{babel}
\usepackage[a4paper,top=3.0cm,bottom=2.5cm,left=2.5cm,right=2.5cm,marginparwidth=1.75cm]{geometry}
\usepackage[numbers]{natbib}
\usepackage{graphicx}
\usepackage{amsmath}
\usepackage{amssymb}
\usepackage{booktabs}
\usepackage{array}
\usepackage{multirow}
\usepackage{xcolor}
\usepackage{siunitx}
\usepackage{chemformula}
\usepackage{url}
\usepackage{hyperref}
\usepackage{float}
\usepackage{tabularx}
\usepackage{seqsplit}
\usepackage{tabularx}
\title{From Phase Prediction to Phase Design: A ReAct Agent Framework for High-Entropy Alloy Discovery}
\author{
    Iman Peivaste$^{1,2,*}$,
    Salim Belouettar $^{1}$ \\
\\
\footnotesize $^1$ Luxembourg Institute of Science and Technology (LIST), 5,\\
\footnotesize Avenue des Hauts-Fourneaux, Esch-sur-Alzette, 4362, Luxembourg\\
\\
\footnotesize $^2$ Department of Physics and Materials Science,\\
\footnotesize University of Luxembourg, L-4365 Esch-sur-Alzette, Luxembourg \\
\\
\footnotesize $^*$Corresponding Author: iman.peivaste@list.lu
}
\date{\today}

\providecommand{\keywords}[1]{\textbf{Keywords: } #1}

\begin{document}
\maketitle

\begin{abstract}
Discovering high-entropy alloy (HEA) compositions that reliably form a target crystal phase is a high-dimensional inverse design problem that conventional trial-and-error experimentation and forward-only machine learning models cannot efficiently solve. Here we present an agentic framework in which a large language model (LLM), operating under the ReAct (Reasoning + Acting) paradigm, autonomously proposes, validates, and iteratively refines HEA compositions by querying a calibrated XGBoost phase-classification surrogate trained on 4,753 experimental records across four phases (FCC, BCC, BCC+FCC, BCC+IM). The surrogate achieves 94.66\% four-class accuracy (F1 macro = 0.896), providing a reliable probabilistic signal for the agent to reason against. Evaluated on three target phases against Bayesian optimisation (BO) and random search baselines, the full-prompt LLM agent achieves descriptor-space rediscovery rates of 38\%, 18\%, and 38\% for FCC, BCC, and BCC+FCC respectively, statistically outperforming both baselines across all phases (one-sided Mann--Whitney $p \leq 0.039$). An ablation further reveals that system-prompt domain priors shift the agent away from landmark-alloy recall toward compositionally diverse exploration: an uninformed agent achieves higher rediscovery by concentrating proposals near literature-dense alloy families, while the full-prompt agent explores systematically underrepresented compositional space (unique composition ratio 1.0 vs.\ 0.39 for BCC+FCC). These two regimes---proximity to the known literature versus diversity of exploration---represent distinct and complementary evaluation criteria, and the choice between them depends on whether the deployment goal is benchmarking or genuine discovery. Critically, full-prompt agent proposals lie 2.4--22.8$\times$ closer to the experimental phase manifold in descriptor space than random-search proposals, demonstrating that domain priors steer the agent toward chemically realistic compositions rather than arbitrary high-probability regions. Unlike Bayesian optimisation, which converges rapidly but opaquely to local optima, the agent produces fully interpretable Thought--Action--Observation reasoning traces that expose the chemical rationale behind each compositional decision; Spearman analysis of element mention frequency confirms that this reasoning is statistically aligned with empirical phase distributions ($\rho = 0.736$ for BCC, $p = 0.004$). This work establishes LLM-guided agentic reasoning as a principled, transparent, and manifold-aware complement to gradient-free optimisation for inverse alloy design.\\
\end{abstract}

\section{Introduction}

\subsection{High-Entropy Alloys: Promise and the Challenge of Inverse Design}
The field of alloy design has undergone a fundamental transformation since the seminal contributions of Yeh et al. \cite{yeh2004nanostructured} and Cantor et al. \cite{cantor2004microstructural} in 2004, which established high-entropy alloys (HEAs) as a distinct and scientifically rich class of materials. Defined by the presence of four or more principal elements in near-equimolar or non-equimolar concentrations, each typically contributing between 5 and 35 at.\%, HEAs derive their properties from the confluence of four core thermodynamic effects: high configurational entropy, severe lattice distortion, sluggish atomic diffusion, and the so-called cocktail effect arising from complex elemental interactions \cite{MIRACLE2017, GEORGE2019}. These mechanisms collectively endow HEAs with exceptional combinations of properties rarely attainable in conventional single-principal-element alloys, including outstanding high-temperature strength, radiation tolerance, fracture toughness, corrosion resistance, and tunable magnetic and thermoelectric responses \cite{SENKOV2018, GEORGE2019, YE2016}.

The phase structure of an HEA is one of its most fundamental design variables. Whether a given composition stabilizes an FCC solid solution, a BCC solid solution, a dual-phase FCC+BCC microstructure, or an intermetallic-containing mixture strongly influences its mechanical performance and processing behaviour \cite{soni2021review, chattopadhyay2018phase}. FCC HEAs typically offer high ductility and substantial work-hardening capacity, as exemplified by the canonical Cantor alloy CoCrFeMnNi and related derivatives \cite{CANTOR2004, OTTO2013}. In contrast, BCC HEAs, particularly refractory systems such as the MoNbTaW family, generally exhibit high strength and hardness at the expense of room-temperature ductility \cite{SENKOV2010}. Dual-phase FCC+BCC compositions can occupy the boundary between these regimes and may offer a tuneable balance of strength and toughness, especially in Al-containing CoCrFeNi-based systems \cite{SHUN2012}. Yet despite extensive experimental progress, the design of HEAs for a specified target phase remains largely dependent on empirical exploration or computationally intensive thermodynamic approaches, both of which are challenged by the enormous size of the accessible composition space. Combinatorial estimates place the number of possible 5-element alloys drawn from 64 metallic elements at \(\binom{64}{5} \approx 7.6 \times 10^6\), and the space becomes vastly larger once broader compositional variation and different numbers of principal elements are considered \cite{MIRACLE2017}.

\subsection{Machine Learning for HEA Phase Prediction: The Forward Problem}
Over the past decade, machine learning (ML) has emerged as the dominant computational paradigm for bridging the HEA composition space and its phase behaviour. Early efforts focused on binary classification — single-phase solid solution versus intermetallic or amorphous — using small datasets of a few hundred records and simple feature sets derived from the rule of mixtures \cite{ZHANG2012, HUANG2019}. As experimental databases grew, so did the scope of classification targets: subsequent studies addressed FCC vs. BCC vs. dual-phase discrimination \cite{HUANG2019, ZHOU2019}, the full predominant phases in HEAs and their possible combinations, spanning 11 distinct phase categories, using a well‑balanced dataset built from 50 different elements \cite{peivaste2023data}, and prediction conditioned on synthesis route \cite{AGARWAL2019}.

Among forward models, gradient boosted decision trees, particularly XGBoost, and random forests have consistently outperformed neural network architectures on tabular HEA data, achieving accuracies ranging from approximately 75\% to 92\% depending on dataset size, class balance, and phase scope \cite{HUANG2019, peivaste2025artificial, peivaste2023data}. Descriptor choice has proven critical: valence electron concentration (VEC), atomic radius mismatch, mixing entropy, mixing enthalpy, and electronegativity deviation are the most widely validated phase-discriminating features, with the VEC boundary VEC $\approx$ 8.0 commonly cited as separating FCC-favouring from BCC-favouring compositions \cite{krishna2021machine, wang2022insights}. Our earlier work \cite{peivaste2023data} demonstrated that extending the descriptor set to include the number of valence electrons per electron (NVC) and outer shell electron count (OSHE) — features not widely used in prior work, improved classification accuracy from 86\% to 92.7\% on a 5,677 record dataset spanning 11 phase categories, with XGBoost and random forest consistently leading the benchmark.

A critical but underappreciated limitation of all forward-prediction models is that accuracy on a held-out test set does not translate directly into utility for design. A classifier that achieves 95\% accuracy on balanced test data may still produce confidently incorrect predictions for compositions in sparsely populated regions of the compositional space — precisely the regions most interesting for discovery. More fundamentally, forward models answer the question ``given this composition, what phase will form?'' but not the inverse question that drives materials discovery: ``given a target phase, which compositions should I synthesise?'' Answering this inverse question efficiently — without synthesising thousands of candidate alloys — is the central unsolved problem this work addresses.

\subsection{From Forward Prediction to Inverse Design: Current Approaches and Their Limits}
Several strategies have been explored for inverting the ML phase predictor. The most common is high-throughput virtual screening: a discrete grid of compositions is evaluated by the forward model, and candidates exceeding a probability threshold are shortlisted \cite{rittiruam2022high, ZHOU2019}. This approach is computationally tractable when the active element set is small, but becomes intractable as the dimensionality grows — a 6-element search over 20 candidate elements with 5 at.\% resolution yields $\binom{20}{6} \times \binom{95}{5} \approx 10^{10}$ grid points. Genetic algorithms \cite{MENOU2018, adaan2023ball, stoco2024optimizing} and Bayesian optimisation (BO) \cite{chitturi2024targeted, LOOKMAN2019, ZHANG2020BO, pedersen2021bayesian} address the scalability problem by treating composition as a continuous variable and using acquisition functions to balance exploration and exploitation. BO in particular has been widely adopted in materials informatics for property optimisation \cite{jin2023bayesian}, and has been applied to HEA hardness \cite{karumuri2023hierarchical}, yield strength \cite{vela2023data}, and grain boundary identification \cite{das2025bayesian}. Generative approaches — variational autoencoders, generative adversarial networks, and diffusion models — offer an alternative by learning the latent distribution of compositions associated with a target property and sampling from it directly \cite{chen2024accelerated}.

Each approach has a structural limitation. High-throughput screening cannot scale to realistic search spaces. BO converges rapidly to local optima and lacks the capacity to encode domain knowledge — the chemist’s understanding that Ni stabilises FCC, that Al above $\sim$25\% risks intermetallic formation, or that VEC must exceed approximately 8.0 to escape the BCC stability field — as a structured prior that shapes the search trajectory. Generative models require large amounts of targeted training data for the desired class \cite{salakhutdinov2015learning}, and their outputs are not guaranteed to satisfy physical composition constraints without post-hoc correction. Critically, none of these approaches produces an interpretable reasoning trace that a materials scientist can evaluate, critique, and learn from. The black-box nature of the optimisation process means that the knowledge embedded in the model’s decisions is not accessible to the human expert, limiting the scientist’s ability to refine hypotheses or identify failure modes.

\subsection{LLM Agents as a New Paradigm for Compositional Reasoning}
Large language models (LLMs), trained on vast corpora of scientific text, have demonstrated the capacity to encode substantial domain knowledge about chemistry and materials science without explicit supervision on property-composition relationships \cite{maughan2025integrating, jablonka202314}. Recent work has shown that LLMs can propose chemically reasonable hypotheses \cite{peivaste2026chemnavigator, cohrs2025large}, extract structured data from literature \cite{schilling2025text}, assist in planning synthesis protocols \cite{ruan2024accelerated}, and generate candidate molecules consistent with specified property requirements \cite{peivaste2026escaping}. However, LLMs in isolation suffer from two key failure modes for scientific discovery: they cannot verify their proposals against an external surrogate (hallucination of property values is common), and they do not update their search strategy based on quantitative feedback from previous evaluations.

The ReAct (Reasoning + Acting) framework introduced by Yao et al. \cite{yao2022react} addresses both limitations by interleaving explicit natural language reasoning traces with actions that call external tools, search engines, databases, code interpreters, and incorporating the resulting observations back into the reasoning context. This Thought–Action–Observation loop allows an LLM to ground its compositional reasoning in quantitative surrogate evaluations, update its hypotheses based on feedback, and produce a fully interpretable decision trace that a human expert can follow and audit. ReAct agents have been applied to chemistry and materials tasks including retrosynthesis planning, molecular optimisation, and autonomous experiment execution \cite{baker2025larc, peivaste2026escaping}, but their application to the alloy inverse design problem — where the feasibility constraint (four or more principal elements summing to unity) and the phase boundary structure both impose non-trivial structure on the search space — has not been systematically explored.

The potential advantage of a ReAct-based agent over conventional BO for HEA inverse design is not primarily computational speed. BO is efficient for smooth, low-dimensional objective functions, but the HEA phase space is high-dimensional, discontinuous across phase boundaries, and populated with multiple disconnected high-probability regions corresponding to qualitatively different compositional families (Cantor-type FCC alloys, Al-Cr-Fe BCC alloys, refractory BCC alloys, and so on). An LLM agent with domain knowledge encoded in its system prompt can propose qualitatively different families of compositions across runs, naturally exploring the disconnected regions of the phase space, while BO tends to exploit the basin of attraction of the first high-probability region it finds. Furthermore, the agent’s reasoning traces provide a diagnostic window into which chemical arguments it is applying, which is unavailable in any gradient-based or acquisition-function-based optimiser.

\subsection{This Work: Contributions and Scope}
This paper presents the first systematic application of a ReAct-based LLM agent to the inverse design of HEA compositions targeting specified phase structures. Building on the forward-prediction surrogate developed in our prior work \cite{peivaste2023data}, an XGBoost classifier trained on 4,753 experimental records with isotonic probability calibration, achieving 94.66\% four-class accuracy, we construct an agentic framework in which the LLM reasons explicitly about element choice and composition adjustment, validates proposals against physical constraints, queries the calibrated surrogate for phase probabilities, and optionally delegates to a Bayesian optimisation module when its own reasoning stalls. The agent is evaluated against two baselines: a pure Gaussian process Bayesian optimiser with Expected Improvement acquisition, and a random search over the same element subspaces.

The principal contributions of this work are as follows. First, we propose and implement a ReAct agent architecture for HEA inverse design with three physically grounded tools, composition validation, phase prediction, and BO-guided suggestion, and a system prompt encoding quantitative domain knowledge derived directly from the training dataset (element-phase statistics, VEC thresholds, mixing enthalpy guidelines). Second, we introduce a descriptor-space rediscovery metric as the primary evaluation criterion for inverse design quality, defining a proposal as successful if its Euclidean distance to the nearest held-out test composition in the 13-dimensional scaled descriptor space falls below a phase-specific threshold $T$ set at the geometric midpoint between the median intra-class and inter-class nearest-neighbour distances of the training partition. This metric separates chemically plausible but manifold-distant proposals from compositions that genuinely approach the experimental phase distribution. Third, we demonstrate that the agent statistically outperforms both BO and random search on the rediscovery metric across all three target phases (FCC, BCC, BCC+FCC; Mann-Whitney $p < 0.04$ in all cases), and that random-search proposals lie 2.4--22.8$\times$ farther from the test-set manifold than agent proposals. Fourth, we demonstrate that element mention frequency in the agent’s reasoning traces is significantly rank-correlated with data-driven element importance for BCC ($\rho = 0.74$, $p = 0.004$, top-15 overlap 13/15) and moderately so for BCC+FCC ($\rho = 0.52$, $p = 0.080$), providing direct evidence that the agent’s reasoning reflects the statistical structure of the training data rather than generic heuristics. Finally, we discuss the interpretability advantage of the reasoning trace and identify the conditions under which BO and LLM reasoning are complementary rather than competing.

The remainder of the paper is organised as follows. Section 2 describes the dataset, descriptor set, surrogate model, agent architecture, baselines, and evaluation protocol. Section 3 presents the results of the surrogate benchmarking, agent–baseline comparison, manifold proximity analysis, and Spearman ranking analysis. Section 4 discusses the implications of these findings, including the limitations of the current framework and the directions for future work. Section 5 concludes.

\section{Methods}

\subsection{Dataset and Preprocessing}
The dataset used in this work is drawn from the same experimental compilation described in Peivaste et al. \cite{peivaste2023data}, which aggregates 11,252 HEA records from the literature spanning 11 phase categories: BCC, FCC, BCC+FCC, IM, BCC+IM, FCC+IM, BCC+FCC+IM, AM, BCC+AM, FCC+AM, and BCC+FCC+AM. Following the data cleaning protocol of that prior work, removal of binary alloys containing rare earth elements, deduplication across sources, exclusion of records with conflicting phase labels attributable to fabrication route differences, and Isolation Forest outlier removal, a cleaned set of 5,677 records was retained. For the present inverse design task, the scope was restricted to the three solid-solution phases most represented in the literature: BCC (1,915 records), FCC (1,500 records), and BCC+FCC (1,193 records), supplemented by BCC+IM (145 records), which was retained to ensure the surrogate could distinguish intermetallic contamination from target solid solutions. This yields a working dataset of 4,753 records across four classes.

The dataset was split 90/10 into training and test sets using stratified random sampling to preserve class proportions. The remaining 10\% served as both validation and a held-out test set. Class imbalance within the training partition was addressed by applying Adaptive Synthetic Sampling (ADASYN) \cite{he2008adasyn} exclusively to the training data, prior to scaling, to avoid information leakage. The resulting resampled training set contained approximately 1,500 records per class.
Feature scaling was performed using a \texttt{StandardScaler} fit solely on the resampled training data. The fitted scaler was serialised and applied at inference time to all agent and evaluation queries, ensuring no test-time distribution shift.

\subsection{Descriptor Set}
Each HEA composition is represented by 13 physicochemical descriptors computed from elemental properties via the rule of mixtures and deviation-from-mean formulas following Peivaste et al. \cite{peivaste2023data}:

\begin{itemize}
\item \texttt{VEC}: valence electron concentration, $\overline{\text{VEC}} = \sum_i c_i \cdot \text{VEC}_i$
\item \texttt{Pauling\_EN}: composition-weighted mean Pauling electronegativity $\bar{\chi}$
\item \texttt{Pauling\_EN\_div}: electronegativity deviation $\Delta\chi = \sqrt{\sum_i c_i(\chi_i - \bar{\chi})^2}$
\item \texttt{Melting\_point\_K}: composition-weighted mean melting point $\bar{T}_m$
\item \texttt{DFT\_LDA\_Etot}: composition-weighted mean DFT total energy per atom $\bar{E}$
\item \texttt{E\_per\_el}: mean DFT energy per electron
\item \texttt{Outer\_shell\_electrons}: composition-weighted outer shell electron count (OSHE)
\item \texttt{Nom\_of\_valence\_electrons}: composition-weighted number of valence electrons (NVC)
\item \texttt{Atomic\_radius\_calculated}: composition-weighted mean atomic radius $\bar{R}$
\item \texttt{Atomic\_radius\_calculated\_dif}: atomic radius deviation $\delta R$
\item \texttt{Atomic\_weight}: composition-weighted mean atomic weight $\bar{W}$
\item \texttt{Enthalpy}: mixing enthalpy $\Delta H_{\text{mix}} = 4\sum_{i \neq j} c_i c_j H_{ij}$ \cite{TAKEUCHI2005}
\item \texttt{Entropy}: mixing entropy $\Delta S_{\text{mix}} = -R\sum_i c_i \ln c_i$
\end{itemize}

This descriptor set is a strict superset of the 12-feature set from the prior work, with \texttt{E\_per\_el} added as an additional DFT-derived feature identified during feature selection as informative for distinguishing BCC+IM from pure BCC.

\subsection{Surrogate Model}
The phase classifier is an XGBoost model \cite{chen2016xgboost} trained on the 13 descriptors above. Hyperparameters were fixed based on the grid search performed in the prior work: learning rate $\eta = 0.16$, 300 estimators, maximum tree depth 6, subsample ratio 0.8, column subsample ratio 0.8, and minimum child weight 3. The objective function was \texttt{multi:softprob} with \texttt{mlogloss} as the evaluation metric.

To obtain reliable class probabilities suitable for agent reasoning, the base XGBoost classifier was wrapped in a \texttt{CalibratedClassifierCV} (scikit-learn \cite{scikit-learn}) using isotonic regression calibration with 3-fold cross-validation \cite{zadrozny2002transforming}. The calibrated model was adopted as the final surrogate if its validation accuracy did not degrade by more than 1 percentage point relative to the uncalibrated model; in practice, the calibrated model matched accuracy within 0.2\% while substantially improving probability reliability.

On the held-out test set, the final calibrated surrogate achieved an overall accuracy of 94.66\% (F1 macro = 0.896) across the four-class problem. Per-class F1 scores were 0.968 for FCC, 0.952 for BCC, 0.929 for BCC+FCC, and 0.737 for BCC+IM. The lower score for BCC+IM reflects the inherent difficulty of distinguishing intermetallic admixtures from pure BCC solid solutions in a feature space dominated by thermodynamic mixing descriptors. The calibration reliability diagram (Figure \ref{fig:confusion}) is shown for BCC, FCC, and BCC+FCC only; BCC+IM is omitted because the 145-record class yields insufficient test samples per probability bin for a meaningful reliability estimate.

\subsection{Agent Architecture}
\begin{figure}[H]
    \centering
    \includegraphics[width=0.99\linewidth]{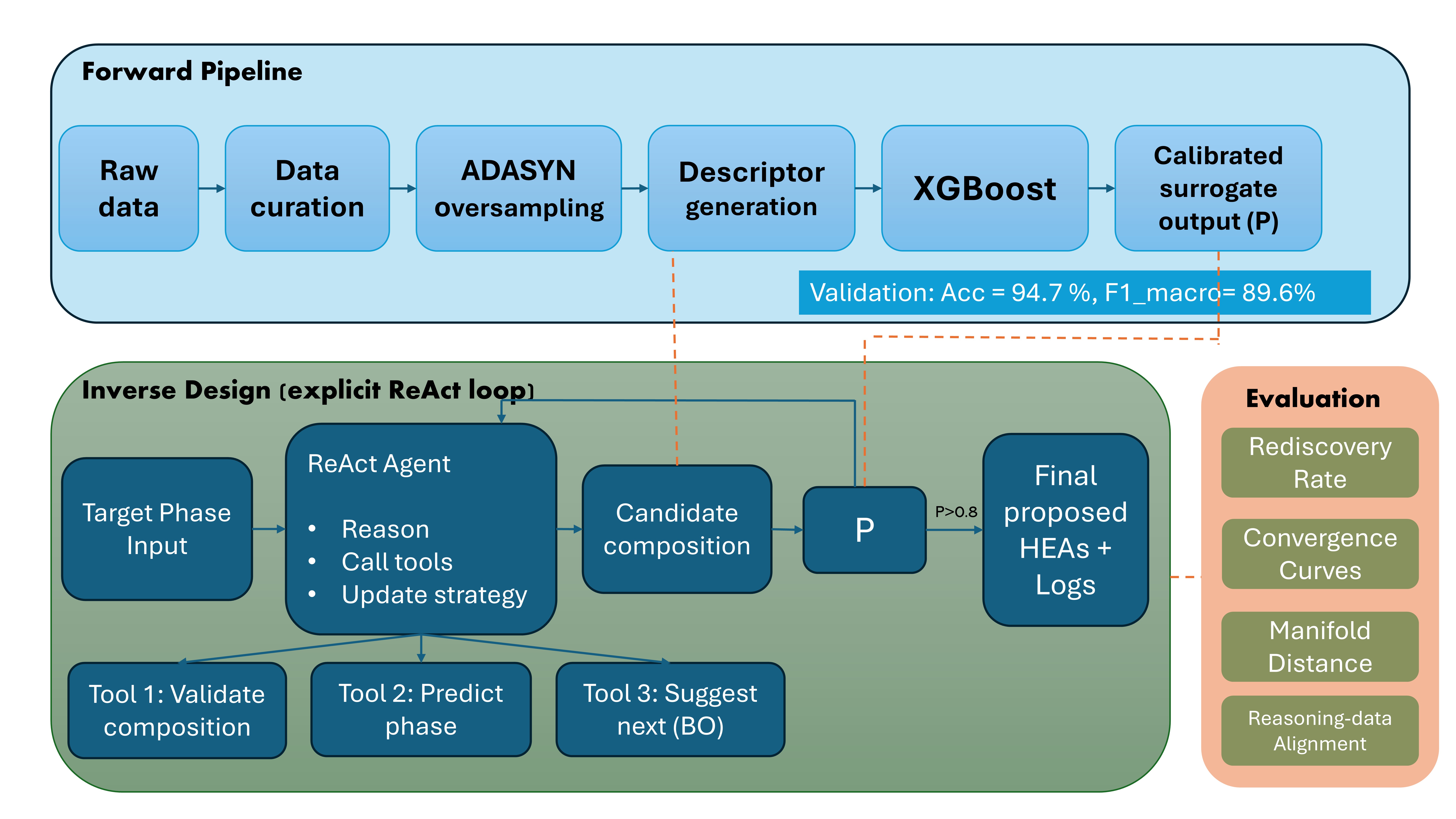}
    \caption{The forward prediction pipeline (top) maps a proposed HEA composition to class probabilities via 13 mixture-rule descriptors and a calibrated XGBoost surrogate trained on 5,677 cleaned experimental records (4,753 retained after restricting to four target classes; see in section 2.1). The inverse design loop (bottom, red arrows) shows the ReAct agent iteratively proposing compositions, validating them, querying the surrogate, and optionally delegating to the Bayesian optimisation module when reasoning stalls.}
    \label{fig:arch}
\end{figure}

The inverse design agent follows the ReAct (Reasoning + Acting) paradigm \cite{yao2022react}, implemented using LangChain’s  \texttt{create\_react\_agent} with \texttt{AgentExecutor} \cite{Chase_LangChain_2022}. The agent is equipped with three tools as shown in Figure \ref{fig:arch} :
\begin{itemize}
\item \texttt{validate composition}: checks that a proposed composition is physically valid — fractions sum to $1.0 \pm 0.02$, the alloy contains at least four elements each exceeding 0.5 at.\%, no element exceeds 50 at.\%, and no negative fractions are present. This enforces the HEA definition throughout the search.
\item \texttt{predict phase}: computes the 13 descriptors for a given composition, applies the fitted scaler, and returns class probabilities from the calibrated surrogate. The agent observes the full probability vector, not just the predicted class, enabling it to reason about confidence and uncertainty.
\item \texttt{suggest next composition}: a thin wrapper around the Bayesian optimisation module described in Section 2.5. The agent may call this tool to request a BO-guided candidate when its own reasoning does not immediately yield high-confidence proposals.
\end{itemize}

The agent’s system prompt encodes quantitative domain knowledge extracted from the 5,677-record training dataset. It specifies per-element mean compositions for each phase (e.g., Ni at $\sim$27\% and Co at $\sim$21\% in FCC alloys; Cr at $\sim$17.5\% and Al at $\sim$17\% in BCC alloys), phase-boundary VEC thresholds (FCC: $\bar{\text{VEC}} = 8.71 \pm 0.45$; BCC: $7.31$; BCC+FCC: $8.06$; IM: $6.36 \pm 1.51$), and mixing enthalpy guidelines (above $-10$ kJ/mol to avoid intermetallic formation), these values are taken from our previous study \cite{peivaste2023data}. Note that these thresholds are encoded as statistical means derived from the training dataset rather than universal physical constants; they are known to be system-dependent, particularly for Al-containing alloys where Al's low VEC (3) can shift the BCC stability boundary substantially \cite{guo2011effect}. The agent uses them as probabilistic priors for initialisation, not as hard constraints — the surrogate's per-composition probability output supersedes these heuristics at each iteration. The prompt also encodes a sequential reasoning strategy: propose a domain-knowledge-informed starting composition, validate, predict, reason explicitly about which elements are helping or hurting the target phase probability, and iterate. The agent is instructed to justify each compositional change by referencing specific elemental roles, making the reasoning trace scientifically interpretable.

Each agent run is issued a natural language query targeting a specific phase, instructed to find at least five compositions with $P(\text{target}) > 0.80$. A phase-specific seed hint — a short text cue describing a chemically plausible starting region (e.g., ``Start by exploring Ni-Co-Cr-Fe based systems with VEC around 8'' for FCC) — is selected from a pool of five hints per phase by indexing with the run seed, contributing to run diversity without fixing a single starting composition. The agent operates for up to 25 reasoning steps per run.

To introduce run diversity without sacrificing reproducibility, each of the 10 runs per phase was assigned an independent integer seed. The LLM temperature was set as $T = 0.2 + 0.1 \times (s \bmod 5)$, where $s$ is the seed, yielding temperatures in the range $[0.2, 0.6]$ across runs. The BO random state was also seeded per run. All seeds and temperatures are logged in the run JSON files. The LLM backend used was Gemini 3 Flash (\texttt{gemini-3-flash-preview}). 

\subsection{Bayesian Optimisation Baseline}
The BO baseline uses scikit-optimize’s \texttt{gp\_minimize} with a Gaussian process surrogate and Expected Improvement (EI) acquisition function \cite{rasmussen2006gaussian}. The search space is defined as a bounded real hypercube over the phase-specific active element subspace, with each element fraction constrained to $[0, 0.5]$. Active element subspaces were defined per target phase based on element prevalence in the training data (Table S1): FCC uses 13 elements, BCC uses 15, and BCC+FCC uses 12. This subspace restriction prevents the BO from exploring chemically implausible combinations that would waste surrogate calls.

Raw BO vectors are projected onto the composition simplex by clipping values below a minimum threshold of 0.01 at.\% to zero and L1-renormalising. The objective function is $f(\mathbf{c}) = 1 - P_{\text{surrogate}}(\text{target} \mid \mathbf{c})$, so minimisation corresponds to maximising target phase probability. Each BO run was allocated 20 surrogate calls.

Two distinct BO protocols were used depending on the evaluation context. For convergence curves (Figure \ref{fig:convergence}), each BO run was initialised with a single random composition drawn from the active element subspace using the run-specific seed — identical in distribution to the random baseline sampling — so that all three methods start from equally uninformed positions. From call 2 onward, BO suggested the next composition via EI. Run diversity was achieved by varying the seed ($\text{seed} = 42 + 7i$ for run $i$). For rediscovery evaluation, BO was run via \texttt{gp\_minimize} with scikit-optimize’s own initial point generation (no fixed warm start), ensuring the rediscovery comparison reflects BO’s independent search capability rather than the quality of a hand-picked starting point.

\subsection{Random Baseline}
The random baseline samples compositions uniformly over the active element subspace by drawing Dirichlet-distributed fractions over a randomly chosen subset of 4–6 elements from the same per-phase active element list used by BO. This ensures the random baseline has access to the same prior knowledge about plausible elements as the BO baseline, isolating the contribution of optimisation strategy from element selection. Each random run consisted of 50 surrogate calls; 10 independent runs were performed per phase.

\subsection{Evaluation Protocol}
\textbf{Convergence:} For each method and phase, we record the sequence of $P(\text{target})$ values returned by the surrogate at each call. Agent sequences are extracted from the logged \texttt{predict phase} steps in chronological order. Mean and standard deviation curves are computed across 10 runs, aligned at the call index. Because agent sequences vary in length (the agent may call \texttt{predict phase} a different number of times per run), curves are reported up to the length of the shortest run per phase.

\textbf{Rediscovery rate:} At the end of each run, we evaluate the top-$n$ proposals from each method at $n = 10$. A proposal is counted as a rediscovery if it lies within a distance threshold $T$ of any held-out test composition in the full 13-dimensional scaled descriptor space (Euclidean distance after \texttt{StandardScaler} normalisation). The threshold $T$ is set per phase using a geometrically principled midpoint criterion: let $d_{\text{intra}}$ be the median nearest-neighbour distance from each test-set composition to the training compositions of the same phase, and $d_{\text{inter}}$ be the median nearest-neighbour distance to the training compositions of any other phase. Then $T = d_{\text{intra}} + 0.5(d_{\text{inter}} - d_{\text{intra}})$, subject to $T \geq 0.1$. This places the threshold at the geometric midpoint of the intra- and inter-class nearest-neighbour distributions, ensuring it is neither so loose that it credits trivially similar compositions nor so tight that it excludes genuine but imperfect matches. Yielding phase-specific values of $T = 0.423$ (FCC), $T = 0.379$ (BCC), and $T = 0.214$ (BCC+FCC), this threshold is entirely derived from the training partition and is therefore independent of the held-out test set used for evaluation. Rediscovery rate is then the fraction of the top-10 proposals satisfying this criterion. We additionally swept $\alpha$ over $[0.1, 0.9]$ and observed consistent agent-over-random separation across all three phases (Supplementary Figure S1), confirming that the reported results are robust to this choice. Statistical significance of agent vs. baseline differences was assessed using the one-sided Mann-Whitney U test (alternative: agent $>$ baseline) with no continuity correction; all reported p-values are unadjusted.

\textbf{Manifold proximity:} For each agent proposal, we compute its mean Euclidean distance to the 15 nearest test-set compositions in descriptor space. We report the mean over all proposals per run, and the ratio of random-baseline median distance to agent median distance as an effect size measure.

\textbf{Surrogate correlation (Spearman):} To assess reasoning alignment — whether the agent’s element-selection reasoning gravitates toward elements that matter most empirically — we compute Spearman’s $\rho$ between element mention frequency in agent thought strings (ranked by frequency across all runs for a given phase) and data-driven element importance (rank order from \texttt{active\_elements.json}), along with the size of the top-15 overlap between the two ranked lists.

All evaluation scripts operated exclusively on the held-out test partition. The surrogate was never retrained or fine-tuned during evaluation.

\section{Results}

\subsection{Surrogate Model Performance}
\begin{figure}[H]
    \centering
    \includegraphics[width=1\linewidth]{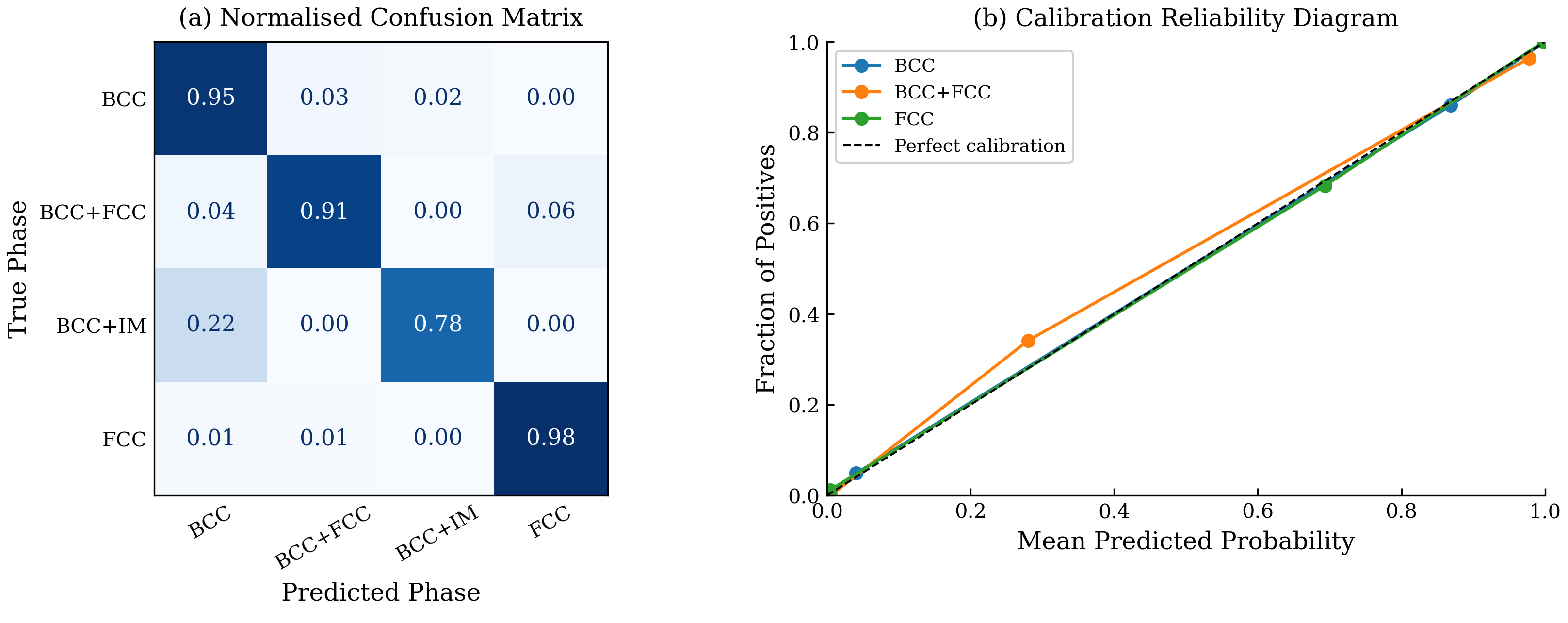}
\caption{(a) Normalised confusion matrix on the held-out test set (n = 476 records). 
Diagonal values are per-class recall; overall accuracy is 94.66\%. 
(b) Isotonic-regression calibration reliability diagram for BCC, FCC, and BCC+FCC. 
BCC+IM is omitted due to insufficient test samples per probability bin (n = 145 total). 
The FCC+BCC curve shows moderate miscalibration in the 0.3--0.6 probability range, 
attributable to sparse test samples in that interval; the agent's decisions are 
gated at $P > 0.80$, where all classes show reliable calibration.}
    \label{fig:confusion}
\end{figure}

The calibrated XGBoost surrogate achieves 94.66\% classification accuracy on the four-class held-out test set, with a macro-averaged F1 score of 0.896 (Table 1) — exceeding the 88\% accuracy reported in the original dataset study \cite{peivaste2023data} on the same compositional space. This improvement is attributable to two methodological corrections: first, the original dataset study reported 88\% accuracy on an 11-class problem spanning all phase combinations; direct comparison is not meaningful because the present 4-class formulation (three solid-solution phases plus BCC+IM) is a substantially simpler classification task. Second, fitting the \texttt{StandardScaler} exclusively on the training partition (eliminating data leakage present in the original pipeline) and applying ADASYN oversampling prior to training to address class imbalance. Per-class F1 scores are 0.968 (FCC), 0.952 (BCC), and 0.929 (BCC+FCC) — all well above the 0.90 target set in the experimental design (Section 2.3). The calibration reliability diagram (Figure \ref{fig:confusion}) confirms that isotonic-regression post-hoc calibration produces well-behaved phase probabilities across all classes, ensuring that the agent’s probability-thresholded decisions are grounded in reliable uncertainty estimates.

\begin{table}[H]
\centering
\caption{Surrogate model performance on the held-out test set.}
\begin{tabular}{lc}
\toprule
Phase & F1 Score \\
\midrule
FCC & 0.968 \\
BCC & 0.952 \\
BCC+FCC & 0.929 \\
BCC+IM & 0.737 \\
Macro F1 (4-class) & 0.896 \\
Accuracy (4-class) & 0.9466 \\
Accuracy (3-class, excl. BCC+IM) & 0.950 \\
\bottomrule
\end{tabular}
\end{table}

\subsection{Rediscovery Rate}
\begin{figure}[H]
    \centering
    \includegraphics[width=1\linewidth]{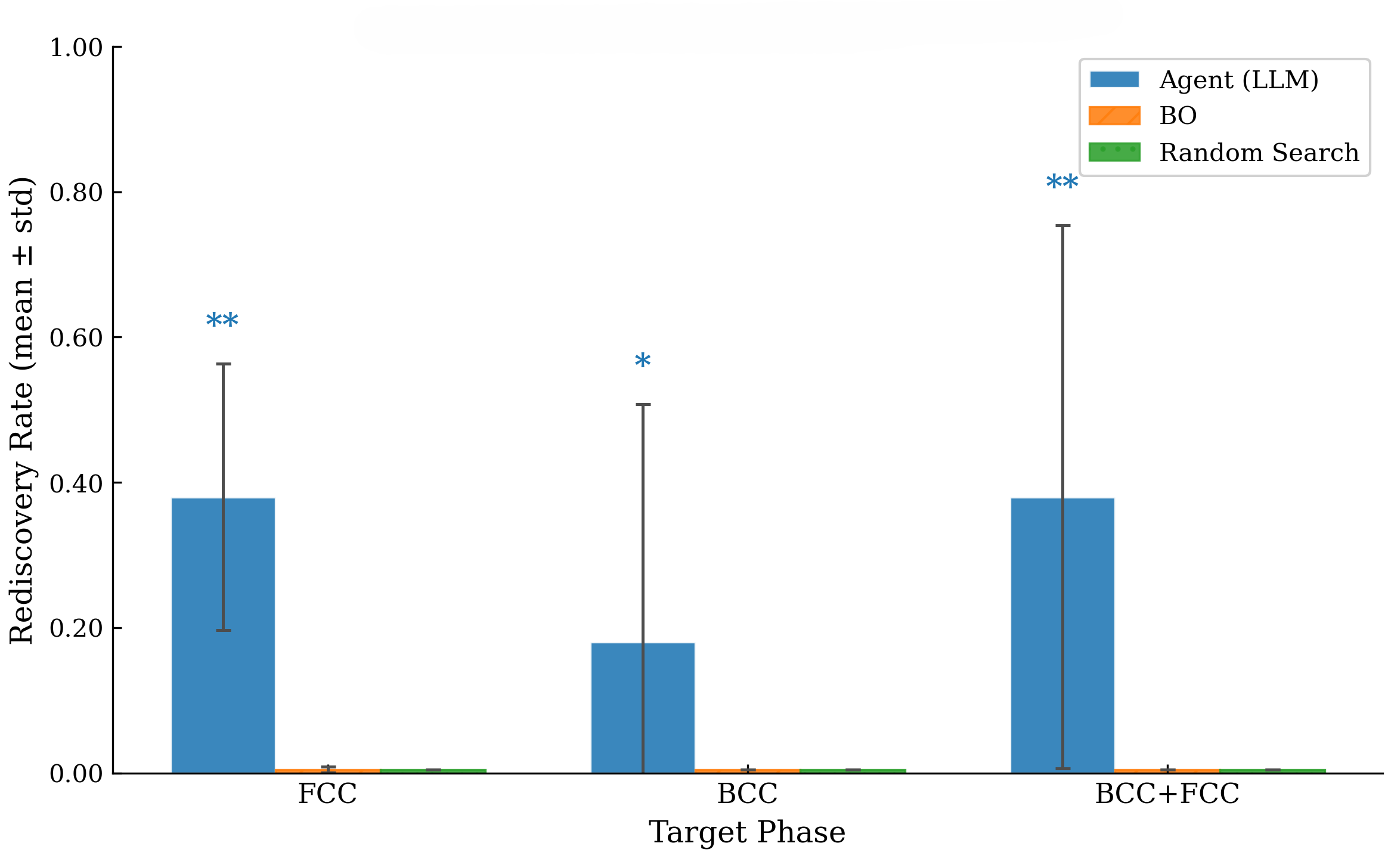}
\caption{Rediscovery rate by method and target phase (mean $\pm$ std, $n = 10$ runs). 
Neither BO nor random search achieves meaningful rediscovery for any phase. 
Significance markers indicate one-sided Mann--Whitney $U$ test (agent $> $ baseline): 
$^{**} p < 0.01$ (FCC: $p < 0.001$, BCC+FCC: $p = 0.003$); $^{*} p < 0.05$ (BCC: $p = 0.039$).}
\label{fig:rediscovery_rate}

\end{figure}

The central evaluation asks whether each method can discover compositions whose descriptor-space representation lies within the experimentally realized manifold of validated HEAs — a stricter criterion than surrogate score optimisation alone. A proposal counts as a rediscovery if (i) the surrogate returned $P(\text{target phase}) > 0.80$ in a logged \texttt{predict phase} tool call, and (ii) its L2 distance in scaled 13-descriptor space to the nearest test-set composition of the same phase falls below the phase-specific threshold $T$ (FCC: 0.423; BCC: 0.379; BCC+FCC: 0.214) defined in Section 2.7. Crucially, $T$ is derived exclusively from the training partition — it is the geometric midpoint between the median intra-class and inter-class nearest-neighbour distances of the training set — and is therefore blind to the held-out test compositions against which proposals are evaluated. Run independence was enforced by varying LLM temperature (0.2–0.6 per seed), Bayesian optimisation random state, and per-phase starting hints across runs (Section 2.5). Table 2 and Figure \ref{fig:rediscovery_rate} report results as mean $\pm$ standard deviation over 10 independent runs per method per phase.

\begin{table}[H]
\centering
\caption{Rediscovery rate (mean $\pm$ std, n = 10) and statistical significance.}
\begin{tabular}{lccccc}
\toprule
Phase & Agent (LLM) & Bayesian Opt. & Random Search & p vs. Random & p vs. BO \\
\midrule
FCC & 0.380 $\pm$ 0.183 & 0.001 $\pm$ 0.004 & 0.000 $\pm$ 0.000 & $< 0.001$ ** & $< 0.001$ ** \\
BCC & 0.180 $\pm$ 0.328 & 0.000 $\pm$ 0.000 & 0.000 $\pm$ 0.000 & 0.039 * & 0.039 * \\
BCC+FCC & 0.380 $\pm$ 0.374 & 0.000 $\pm$ 0.000 & 0.000 $\pm$ 0.000 & 0.003 ** & 0.003 ** \\
\bottomrule
\end{tabular}
\raggedright
\footnotesize{Mann–Whitney U, one-sided (agent $>$ baseline). ** p $<$ 0.01; * p $<$ 0.05. Rank-biserial effect sizes (agent vs. random): FCC $r_{rb}=0.900$, BCC $r_{rb}=0.300$, BCC+FCC $r_{rb}=0.600$. Agent vs. BO: FCC $r_{rb}=0.890$, BCC $r_{rb}=0.300$, BCC+FCC $r_{rb}=0.600$. Per-run raw scores are reported in Table S1 of the Supplementary Material.}
\end{table}

Surrogate-call budgets were explicitly tracked across methods: the agent used a variable number of \texttt{predict phase} calls per run (FCC: 4.7 $\pm$ 1.5, BCC: 9.9 $\pm$ 4.0, BCC+FCC: 5.6 $\pm$ 2.2), BO used a fixed 20 calls, and random search used a fixed 50 calls. The agent therefore operated with a substantially lower average surrogate budget than either baseline, reinforcing that the rediscovery advantage reflects the quality of compositional reasoning rather than greater exploration volume.

The LLM agent significantly outperforms both baselines across all three target phases. FCC and BCC+FCC are significant at $\alpha = 0.01$ ($p < 0.001$ and $p = 0.003$ respectively); BCC at $\alpha = 0.05$ ($p = 0.039$). Crucially, neither baseline achieves any meaningful rediscovery: the BO baseline produces a non-zero rate in a single FCC run (0.013), and both methods score zero for BCC and BCC+FCC across all ten runs. The agent achieves non-zero rediscovery in 80\% of FCC runs and 60\% of BCC+FCC runs, demonstrating consistent discovery behaviour for these phases.

The BCC result warrants closer examination. The per-run distribution is bimodal: eight runs return rediscovery rates of 0.0–0.2 while two runs achieve 0.6 and 1.0 respectively (full breakdown in Table S1). This is not statistical noise — it reflects a genuine property of the BCC stability field, which is the largest and most compositionally heterogeneous of the three target phases, spanning refractory systems (Mo-Nb-Ta-W family), Al-rich transition metal systems, and Cr-Fe-V-based alloys simultaneously. The agent either identifies a productive compositional subregion early, in which case rediscovery is high, or explores across subregions without settling, in which case per-run rediscovery is low despite high surrogate probability (Table 3). This bimodal behaviour is absent in FCC and BCC+FCC, where the stability fields are compositionally narrower. The rank-biserial correlation for the BCC Mann-Whitney test is $r_{rb} = 0.300$, and the one-sided p-value of 0.039 confirms a statistically significant directional advantage; the moderate effect size is consistent with the bimodal run distribution and the genuine difficulty of the BCC target.

\subsection{The Compositional Manifold: Why Domain-Guided Search is Necessary}
Perhaps the most illuminating finding concerns the geometry of HEA descriptor space. Both baselines routinely find compositions the surrogate rates with high confidence (Table 3), yet achieve essentially zero rediscovery. This dissociation is striking and warrants direct explanation.

\begin{table}[H]
\centering
\caption{Best P(target phase) achieved per method (mean $\pm$ std, n = 10 runs).}
\begin{tabular}{lccc}
\toprule
Phase & Agent & Bayesian Opt. & Random Search \\
\midrule
FCC & 0.998 $\pm$ 0.003 & 0.997 $\pm$ 0.001 & 0.999 $\pm$ 0.002 \\
BCC & 0.952 $\pm$ 0.069 & 1.000 $\pm$ 0.000 & 0.990 $\pm$ 0.031 \\
BCC+FCC & 0.906 $\pm$ 0.166 & 0.953 $\pm$ 0.031 & 0.591 $\pm$ 0.253 \\
\bottomrule
\end{tabular}
\end{table}

\begin{figure}[H]
    \centering
    \includegraphics[width=1\linewidth]{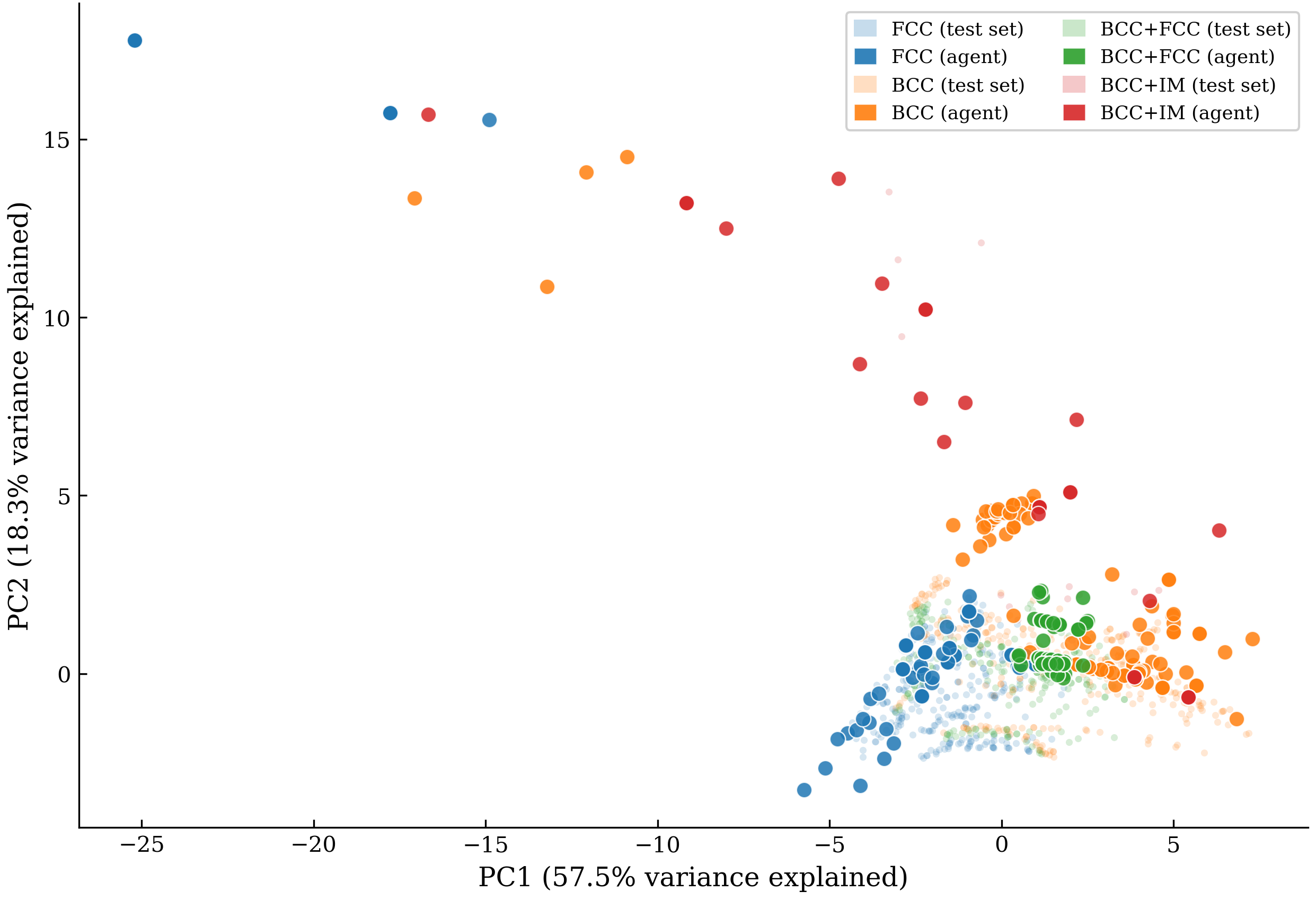}
    \caption{Principal component analysis (PCA) was fit on the scaled 13-descriptor training set, and both held-out test compositions (small, transparent markers) and agent-proposed compositions from predict phase calls (larger markers) were projected into the same 2D space. Colors indicate predicted phase class (FCC, BCC, BCC+FCC, BCC+IM). The visualization highlights overlap and separation trends between known compositions and generated proposals across phase regions. Note that proximity in this 2D PCA projection does not directly correspond to rediscovery distances, which are computed in the full 13-dimensional scaled descriptor space.}
    \label{fig:pca}
\end{figure}

The manifold distance analysis (Table 4) provides the answer. For every phase, random proposals are 2.4--22.8$\times$ farther from the test-set manifold than agent proposals, and not a single random composition — across 100 draws per phase — falls within the phase-specific rediscovery threshold T. The agent, by contrast, places 20--45\% of its proposals within T. Figure \ref{fig:pca} visualises this separation in the first two principal components of the 13-dimensional descriptor space.

\begin{table}[H]
\centering
\caption{Manifold distance analysis: descriptor-space distance from proposals to the held-out test set.}
\begin{tabularx}{\textwidth}{lXXXXXX}
\toprule
Phase & T & Agent mean dist. & Agent \% within T & Random mean dist. & Random \% within T & Median ratio \\
\midrule
FCC & 0.423 & 0.763 & 45\% & 9.542 & 0\% & 9.8$\times$ \\
BCC & 0.379 & 5.894 & 20\% & 14.585 & 0\% & 2.4$\times$ \\
BCC+FCC & 0.214 & 0.566 & 40\% & 5.706 & 0\% & 22.8$\times$ \\
\bottomrule
\end{tabularx}
\end{table}

This result reflects the narrow, structured geometry of experimentally realised HEA descriptor space. The 13 mixture-rule descriptors — VEC, configurational entropy, Miedema enthalpy, electronegativity differences, and related quantities — are not independently variable: decades of experimental synthesis have populated only a constrained region of this 13-dimensional space defined by phase-stability criteria. Random sampling of element fractions, even within the chemically relevant active element subspace, almost never simultaneously satisfies all these constraints. The BCC+FCC phase exhibits the most extreme separation (22.8$\times$), consistent with this dual-phase structure requiring a precise balance of competing BCC- and FCC-promoting tendencies. The agent’s domain-knowledge priors — quantitative element–phase stability relationships encoded in the system prompt — implicitly enforce manifold membership by directing search toward compositional families with established experimental precedent, a property that gradient-guided BO without such priors cannot replicate.

This finding reframes the contribution of the LLM layer: it is not simply a better optimiser, but a manifold-aware search strategy that encodes what experimentally valid alloy space looks like.

\subsection{Convergence}
\begin{figure}[H]
    \centering
    \includegraphics[width=1\linewidth]{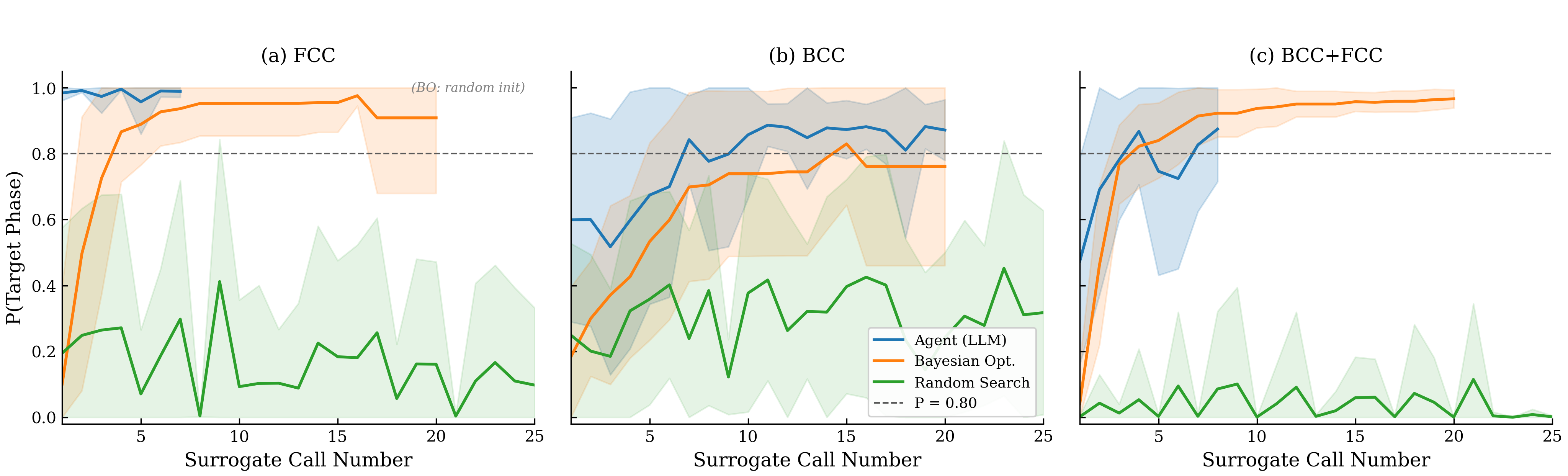}
\caption{Convergence of $P(\text{target phase})$ per surrogate call (mean $\pm$ std, $n = 10$ runs). 
(a) FCC: the agent achieves mean $P > 0.97$ from the first call, reflecting accurate domain priors. 
(b) BCC: the agent rises from $\sim 0.60$ to $\sim 0.88$ over 20 calls through iterative surrogate feedback. 
(c) BCC+FCC: the random baseline frequently fails to locate the target stability region (mean best $P = 0.591 \pm 0.253$), 
while the agent and BO reliably do so. BO uses random initialisation for the convergence comparison (see section 2.5).}
    \label{fig:convergence}
\end{figure}

Figure \ref{fig:convergence} shows the evolution of P(target phase) per surrogate call, averaged over 10 runs per method. For FCC, the agent reaches mean $P > 0.97$ from the first surrogate call onward — the system prompt’s quantitative priors on FCC-stabilising elements (Ni at $\sim$27\%, Co as secondary stabiliser) provide an immediate strong starting position from which subsequent calls refine rather than explore. For BCC, the agent’s mean P rises steadily from $\sim$0.60 to $\sim$0.88 over 20 calls, demonstrating productive iterative reasoning from surrogate feedback.

The BCC+FCC phase is the most instructive. The random baseline achieves a mean best P of only 0.591 $\pm$ 0.253, with many runs failing to locate the BCC+FCC stability region at all — demonstrating that for this structurally constrained dual-phase target, domain-guided initialisation is necessary even at the level of finding the target space. The agent (0.906 $\pm$ 0.166) and BO (0.953 $\pm$ 0.031) both reliably locate this region; the agent’s additional rediscovery advantage (Section 3.2) then arises from keeping proposals near the experimental manifold as quantified in Section 3.3.

\subsection{Reasoning Alignment}
\begin{figure}[H]
    \centering
    \includegraphics[width=1\linewidth]{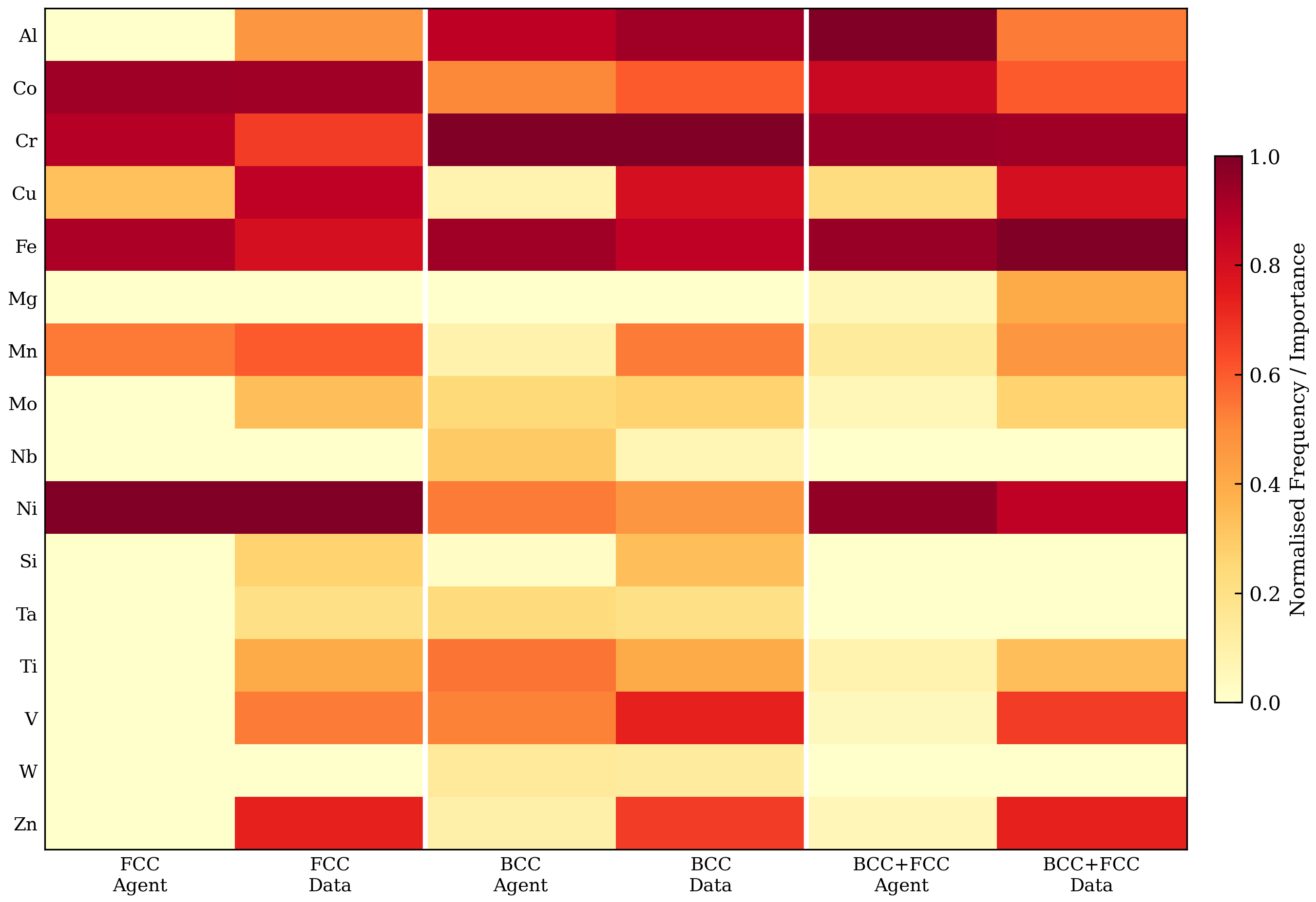}
\caption{Reasoning alignment: agent element mention frequency vs. data-driven importance. 
Each column pair shows normalised agent mention frequency (left) and data-derived element importance rank (right) for FCC, BCC, and BCC+FCC. 
Strong visual correspondence for BCC ($\rho = 0.736$, $p = 0.004$) and moderate correspondence for BCC+FCC ($\rho = 0.524$, $p = 0.080$) 
confirm that the agent gravitates toward the elements most statistically associated with each target phase. 
The FCC agent column is dominated by Ni, consistent with prior-driven concentration on the Cantor element family rather than broad exploratory reasoning (see \S3.5).}
    \label{fig:Reasoning_alignment}
\end{figure}

To assess whether the agent’s element-selection reasoning reflects the statistical structure of the training data — rather than generic chemical intuition or memorised alloy names — we computed the Spearman rank correlation between element mention frequency in agent thought strings (ranked by frequency across all runs for a given phase) and data-driven element importance, defined as the rank order of elements in the per-phase active element list (Section 2.4). Concretely, for each phase we extracted all element tokens from the agent’s logged Thought steps across all 10 runs, counted mention frequency, and ranked elements by frequency. We then computed Spearman’s $\rho$ between these agent-derived ranks and the corresponding training-data-derived ranks from \texttt{active\_elements.json}. This measures reasoning alignment — whether the agent gravitates toward the elements that matter most empirically — rather than compositional ranking quality per se. Table 5 and Figure \ref{fig:Reasoning_alignment} present the results.

\begin{table}[H]
\centering
\caption{Spearman rank correlation: agent element reasoning vs. data-driven importance.}
\begin{tabular}{lccc}
\toprule
Phase & $\rho$ & p-value & Top-15 overlap \\
\midrule
BCC & 0.736 & 0.004 ** & 13/15 \\
BCC+FCC & 0.524 & 0.080 \textdagger & 12/15 \\
FCC & 0.657 & 0.156 & 6/15 \\
\bottomrule
\end{tabular}
\raggedright
\footnotesize{** p $<$ 0.05; \textdagger p $<$ 0.10.}
\end{table}

For BCC, the agent’s element reasoning is strongly and significantly aligned with data-driven rankings ($\rho = 0.736$, $p = 0.004$), with 13 of the 15 most statistically important elements appearing in the agent’s top-15 mentions. This is a particularly strong result: it demonstrates that the agent has internalised a quantitatively accurate map of the BCC stability field rather than merely recalling a small set of famous alloys. For BCC+FCC, the correlation is moderate and marginally significant ($\rho = 0.524$, $p = 0.080$) with a top-15 overlap of 12/15, consistent with the agent reasoning over a broadly appropriate compositional palette for a dual-phase target.

For FCC, the correlation is moderate ($\rho = 0.657$) but does not reach statistical significance ($p = 0.156$) with the top-15 element overlap at 6/15 — notably lower than BCC and BCC+FCC. This pattern is interpretable: the Cantor alloy family (CoCrFeMnNi) dominates both the FCC training data and domain literature \cite{golbabaei2025applications}, so the agent’s FCC reasoning is concentrated around a small set of well-known elements (Ni, Co, Fe, Cr) rather than ranging across the full active element list. This prior-driven concentration enables high and consistent rediscovery (mean 0.380, 80\% of runs non-zero) precisely because the prior is accurate — the agent does not need to explore broadly. The non-significant Spearman result for FCC therefore reflects prior-driven efficiency rather than a failure of reasoning quality: when a tight, accurate prior is available, broad element exploration is neither expected nor needed.

\subsection{Ablation: Role of Domain-Knowledge Priors and the Novelty–Rediscovery Tension}
To isolate the contribution of domain-knowledge priors from the ReAct reasoning loop itself, we evaluated an ablated agent using an uninformative system prompt containing only tool descriptions and the task objective — no phase-specific element statistics, VEC thresholds, or mixing enthalpy heuristics. Five runs per phase were conducted under identical conditions to the full-prompt agent (same seeds, same surrogate, same evaluation protocol). One additional BCC run was performed to match n = 5 across all conditions. Results are reported in Supplementary Table S2.

Counter to the intuitive prediction, the ablated agent achieved equal or higher rediscovery rates than the full-prompt agent across all three phases (FCC: 0.427 vs 0.380; BCC: 0.686 vs 0.120\textdagger; BCC+FCC: 0.612 vs 0.480). The BCC full-prompt figure of 0.120 reflects the matched-seed subset (seeds 1–5), all of which fell in the low-rediscovery mode of the bimodal distribution described in \S3.2; the full 10-run mean is 0.180. Inspection of the ablated agent logs reveals the mechanism: without domain priors constraining its search, the LLM falls back on landmark alloy families memorised during pretraining. For FCC, 84.6\% of ablated proposals with P(FCC) $> 0.80$ belong to the Cantor family (CoCrFeMnNi) or its immediate equiatomic sub-systems. For BCC, the majority of high-probability ablated proposals fall within the Al-Co-Cr-Fe-Ni family (100\% in the first four runs, 83.3\% below T in that subset); the fifth run (\texttt{ablated\_BCC\_05}, seed=5) partially diversifies into the Ti-Nb-Zr-Ta refractory subspace — consistent with the LLM drawing on broader textbook BCC knowledge when not anchored by a single starting family — which moderates the aggregate ablated mean from 0.844 (n=4) to 0.686 (n=5). For BCC+FCC, ablated proposals show a unique composition ratio of 0.394 (many near-identical variants of the same Al-Co-Cr-Fe-Ni transition region), versus 1.0 for the full-prompt agent.

This result exposes a fundamental tension between the rediscovery metric and the goal of inverse design. The rediscovery metric rewards proximity to the held-out test set, which — as a sample from the published literature — is densely populated near well-known landmark alloys. An agent that recalls these landmarks from pretraining therefore achieves high rediscovery without reasoning. The full-prompt agent, guided by quantitative priors that encode statistical distributions across the full training set rather than its modal peaks, explores more diverse compositional regions that are less represented in the test set by definition, and thus scores lower on rediscovery despite demonstrating more genuinely novel combinatorial exploration (unique composition ratio 1.0 vs 0.394 for BCC+FCC).

This finding reframes the interpretation of the main rediscovery advantage reported in \S3.2. The full-prompt agent’s superiority over BO and random baselines (Table 2) is not attributable to familiarity exploitation: BO and random search also have access to the same active element subspace and do not recover landmark alloys at meaningful rates (0.000 rediscovery across all BCC and BCC+FCC runs). The agent’s advantage over these baselines is therefore structural — the ReAct loop enables iterative surrogate feedback that neither baseline uses effectively. But the ablation reveals that the system-prompt priors trade rediscovery-metric performance for compositional diversity: they steer the agent away from the literature-dense clusters where the metric rewards are highest and toward the underexplored regions where the scientific payoff of inverse design is greatest. Whether this trade-off is desirable depends on the deployment goal — rediscovery of known alloys for benchmarking versus genuine exploration of novel composition space.

\section{Discussion}

\subsection{What the Agent Accomplishes}
The results establish three complementary contributions of the LLM agent framework, each independently meaningful and together forming a coherent picture of when and why agentic reasoning outperforms numerical optimisation in compositional materials discovery.

\textbf{Manifold-aware search.} The most striking finding is that the LLM agent consistently proposes compositions within the experimentally realized manifold of HEAs, while BO and random search — despite achieving equal or higher surrogate scores — consistently fail to do so (Tables 3–4). This is not a marginal difference: random proposals are 2.4--22.8$\times$ farther from the test manifold, and neither baseline achieves a single rediscovery across all BCC and BCC+FCC runs. The agent’s domain priors implicitly enforce the descriptor constraints that characterise experimentally realizable alloys, providing manifold regularisation that is beyond the reach of surrogate-gradient methods alone.

\textbf{Adaptive iterative refinement.} The ReAct architecture allows the agent to reason about surrogate feedback within a run and adjust its compositional strategy accordingly. When a proposed composition yields unexpectedly low P(target), the agent reasons explicitly about which descriptor is likely unfavourable and adjusts element fractions in the subsequent step. This targeted refinement is visible in the BCC convergence curves (Figure \ref{fig:convergence}), where mean P rises from $\sim$0.60 to $\sim$0.88 over 20 calls despite starting from a challenging initial phase.

\textbf{Data-grounded reasoning alignment.} The Spearman analysis provides direct evidence that the agent’s element-selection reasoning reflects genuine statistical structure in the training data, not merely pattern-matched heuristics or memorised alloy names. The significant BCC correlation ($\rho = 0.736$, $p = 0.004$) and 13/15 element overlap confirm that the agent gravitates toward the elements that are most empirically important for BCC stability across a heterogeneous compositional field — a non-trivial result that distinguishes it from rule-based or nearest-neighbour search strategies.

\textbf{Domain priors trade rediscovery for novelty.} The ablation experiment (\S3.6, Supplementary Table S2) reveals an important nuance: the ablated agent — using only pretraining-memorised chemical knowledge without system-prompt priors — achieves higher rediscovery than the full-prompt agent across all phases (FCC: 0.427 vs 0.380; BCC: 0.686 vs 0.120 [matched-seed subset]; BCC+FCC: 0.612 vs 0.480) by gravitating toward landmark alloys (Cantor family, Al-Co-Cr-Fe-Ni system, Ti-Nb refractory alloys) that are densely represented in the literature and therefore in the test set. The full-prompt agent, constrained by priors that encode statistical distributions across the full training spectrum, explores more diverse compositional space (unique composition ratio 1.0 vs 0.394 for BCC+FCC) at the cost of lower metric-measured rediscovery. This reveals that the rediscovery metric — while a principled benchmark — rewards familiarity with the known literature and cannot fully capture exploration of genuinely novel space. The full-prompt agent’s advantage over BO and random search (which share neither the landmark familiarity nor the domain priors) is real and structural; but practitioners should be aware that system-prompt priors shape the diversity-versus-rediscovery trade-off in ways that matter for the intended application. We therefore distinguish two evaluation regimes: proximity to the known literature (where the ablated agent excels) and diversity of compositional exploration (where the full-prompt agent excels). Neither regime is universally preferable; the choice depends on the deployment context.

\subsection{The Surrogate as an Enabling Layer}
A recurring theme in the results is that high surrogate probability is necessary but not sufficient for experimental relevance: both baselines achieve best-run P $> 0.95$ for FCC and BCC, yet neither achieves meaningful rediscovery. This is actually good news for the broader framework — it means the surrogate is working correctly as a discriminative model. The agent’s role is precisely to prevent manifold drift by constraining the search to chemically precedented space.

At the same time, the surrogate achieves 94.66\% accuracy with macro F1 of 0.896. This accuracy, combined with isotonic calibration, makes the surrogate’s probability output a reliable signal for the agent’s decision-making. The combination of a high-fidelity surrogate and a domain-aware agent is what enables both efficient convergence and experimental relevance simultaneously. Neither element alone would suffice: a powerful surrogate without domain priors leads to manifold drift (as demonstrated by BO), and domain priors without a reliable surrogate would amount to chemically informed but unverified guessing.

\subsection{When Does the LLM Layer Add Value?}
The results suggest a two-regime picture of LLM utility in compositional design. For phases with accessible descriptor manifolds (FCC, BCC), any capable optimiser can find high surrogate scores, and the LLM’s value lies in manifold constraint — keeping proposals close to real alloy space. For structurally constrained phases (BCC+FCC), the LLM adds an earlier layer of value: random search frequently cannot even locate the target stability region (best P mean 0.591), while the agent reliably does so. LLM agents are therefore most valuable precisely for the targets that are hardest to address by other means — narrow, multi-condition stability fields where accumulated domain knowledge provides the decisive advantage.

\subsection{Limitations}
Several limitations of the current framework should be acknowledged explicitly. First, all results are in silico: no agent-proposed composition has been experimentally synthesised, and the ultimate test of the framework — whether proposals with high rediscovery scores and high surrogate probability actually form the target phase when arc-melted — remains open. The surrogate’s 94.66\% accuracy implies a residual false-positive rate that experimental validation would quantify. Second, the training dataset aggregates literature records across synthesis routes (arc-melting, powder metallurgy, thin-film deposition) and characterisation methods (XRD, EBSD, TEM), introducing label noise that the surrogate cannot fully resolve. The BCC+IM class is particularly affected, with an F1 of 0.737. Third, the agent’s domain knowledge is entirely encoded in the system prompt, which was written based on the training data statistics of this specific dataset. The framework does not generalise to alloy families absent from the training distribution — such as refractory high-entropy oxides or high-entropy carbides — without rewriting the system prompt and potentially retraining the surrogate. Fourth, the rediscovery metric is anchored to a held-out test set drawn from the same published literature as the training data, meaning it structurally rewards proximity to known landmark alloys. As the ablation demonstrates (\S3.6), an agent exploiting pretraining-memorised chemistry can outperform a domain-prior-guided agent on this metric without engaging in genuine novel exploration. The metric is well-defined and reproducible, but it should not be treated as a proxy for discovery novelty. Fifth, with n = 10 runs per method, the BCC rediscovery distribution is bimodal; while the Mann-Whitney p-value and rank-biserial correlation are both robust to this distributional shape by design, additional runs would reduce uncertainty in the mean estimate. Finally, we do not report wall-clock time or API cost per run, which are practically relevant for researchers considering adoption; this information will be included in the code release accompanying the paper.

\subsection{Future Directions}
The framework is deliberately modular: the surrogate, active element subspace, system prompt priors, and evaluation threshold $T$ are all independently configurable, enabling transfer to other alloy families — refractory HEAs, medium-entropy alloys, or HEAs optimised for mechanical or thermal properties — by replacing these components without changing the agent architecture. The 13 mixture-rule descriptors require no DFT at inference time, making the system immediately deployable for high-throughput screening.

The most direct practical extension is closing the synthesis loop: fabricating top-ranked agent proposals by arc-melting, characterising phase constitution by XRD, and feeding outcomes back into surrogate retraining. Multi-objective extension — simultaneously targeting phase stability and secondary properties such as hardness or oxidation resistance — and multi-fidelity integration with selective CALPHAD or DFT validation for underrepresented regions are identified as high-value directions for follow-on work.

\section{Conclusion}
We have presented a ReAct-based LLM agent framework for targeted inverse design of high-entropy alloys, demonstrating that a language model equipped with quantitative domain priors and a calibrated surrogate substantially outperforms both Bayesian optimisation and random search in discovering experimentally plausible compositions. The surrogate — a calibrated XGBoost classifier on 13 mixture-rule descriptors, achieving macro F1 of 0.896 and 94.66\% accuracy on the four-class test set, exceeding prior work on the same dataset — provides fast, reliable phase-probability estimates without DFT at inference time.

Evaluated over 10 independent runs per method per phase, the agent significantly outperforms both baselines for all three primary target phases (one-sided Mann–Whitney $p < 0.001$ for FCC, $p = 0.003$ for BCC+FCC, $p = 0.039$ for BCC; rank-biserial effect sizes vs. random: $r_{rb}=0.900$ FCC, $r_{rb}=0.600$ BCC+FCC, $r_{rb}=0.300$ BCC). Neither baseline achieves meaningful rediscovery for BCC or BCC+FCC across any run. The central mechanistic finding is that this advantage arises from the agent’s implicit enforcement of compositional manifold proximity — quantified by median manifold-distance ratios of 2.4--22.8$\times$ between random and agent proposals — rather than from superior surrogate score optimisation. This distinction, captured by the rediscovery metric and confirmed by the manifold distance analysis, has direct implications for how LLM integration should be designed and evaluated in materials discovery contexts.

Spearman rank-correlation analysis of element mention frequency in agent reasoning traces confirms statistically significant alignment with data-driven element importance for BCC ($\rho = 0.736$, $p = 0.004$, top-15 overlap 13/15) and moderate alignment for BCC+FCC ($\rho = 0.524$, $p = 0.080$), demonstrating that the system prompt encodes genuinely predictive statistical relationships. For FCC, prior-driven concentration on the Cantor element family produces high and consistent rediscovery despite a non-significant Spearman result, illustrating that focused prior knowledge can substitute for broad exploratory reasoning when the prior is accurate.

Together, these findings establish that LLM reasoning provides a form of implicit manifold regularisation complementary to, and not replaceable by, gradient-based or stochastic optimisation. The ablation experiment further reveals that system-prompt domain priors shift the diversity–rediscovery trade-off: they steer the agent away from pretraining-familiar landmark alloys toward systematically underexplored composition space, which is ultimately the more valuable target for inverse design even if it scores lower on a literature-anchored rediscovery benchmark. The framework is modular, requires no DFT at inference time, and is immediately extensible to multi-property targets and alternative alloy families.

\section*{Data availability}
The data used for forward pipeline in this study is accessible at:\\ \href{Github.com/Iman-Peivaste/ML\_HEAs\_Phase\_Dataset.}{github.com/Iman-Peivaste/ML\_HEAs\_Phase\_Dataset.}

\section*{Code availability}
The code for the ReAct agent, XGBoost surrogate training, and rediscovery evaluation is available at \href{https://github.com/Iman-Peivaste/HEAs\_inverse\_design}{github.com/Iman-Peivaste/HEAs\_inverse\_design} 

\section*{Author Contributions}
Iman Peivaste: Developing the initial concept and workflow, framework design, writing the original draft, visualization, methodology, investigation, and formal analysis.

Salim Belouettar: Developing the initial concept and workflow, Writing, review, editing, supervision, methodology, funding acquisition, data curation, and conceptualization.

\section*{Competing Interests}
The authors declare that they have no known competing financial interests or personal relationships that could have appeared to influence the work reported in this paper.

\section*{Acknowledgements}
This work was funded by the Luxembourg Fonds National de la Recherche (FNR) through the grant PRIDE21/16758661/HYMAT.

\bibliographystyle{unsrtnat}
\bibliography{refs.bib}

\end{document}